\documentclass{kluwer}    % Specifies the document style.
\usepackage{epsfig}
\usepackage{graphicx}

\begin{document}
\begin{article}
\begin{opening}
\title{Black Holes on the Brane with Induced Gravity}
\author{G. Kofinas}
\institute{ Department of Physics, University of Crete,\\ 710 03
Heraklion, Crete, Greece \\ }
\author{
 E.
Papantonopoulos and  V. Zamarias}
 \institute{ Department of Physics,
National Technical University of Athens,\\ Zografou Campus GR 157
80, Athens, Greece}

%
%$^{1}$ Department of Physics, University of Crete,\\ 710 03
%Heraklion, Crete, Greece\\ $^{2,\,3}$ Department of Physics,
%National Technical University of Athens,\\ Zografou Campus GR 157
%80, Athens, Greece\\ Leslie \surname{Lamport}}
%\runningauthor{Leslie Lamport} \runningtitle{A Sample Document}

\date{September 30, 2002}

\begin{abstract}
An analysis of a spherically symmetric braneworld configuration is
performed when the intrinsic curvature scalar is included in the
bulk action. In the case when the electric part of the Weyl tensor
is zero, all the exterior solutions are found; one of them is of
the Schwarzschild-$(A)dS_{4}$ form, which is matched to a modified
Oppenheimer-Volkoff interior solution. In the case when the
electric part of the Weyl tensor is non zero, the exterior
Schwarzschild-$(A)dS_{4}$ black hole solution is modified
receiving corrections from the non-local bulk effects. A
non-universal gravitational constant arises, depending on the
density of the considered object and the Newton's law is modified
for small and large distances; however, the conventional limits
are easily obtained.
\end{abstract}
\end{opening}
\section{Introduction}
Branes are solitonic solutions of ten-dimensional string theories.
In the most simplified picture of the braneworld scenario, our
physical world is realized as a four-dimensional hypersurface
embedded in a five-dimensional space called bulk. All matter and
gauge interactions live on the brane, while the gravitational
interactions are effective in the whole five-dimensional space.
This allowed to define a gravitational scale of the whole space,
which if the extra dimensions are large, can be as low as the
$TeV$ scale \cite{dimo,randall}, while the four-dimensional
gravitational scale of our world is at the Planck scale.

The effective brane equations have been obtained \cite{maeda} when
the effective low-energy theory in the bulk is higher-dimensional
gravity. If the dynamics is governed not
 only by the ordinary five-dimensional
Einstein-Hilbert action, but
 also by the four-dimensional Ricci scalar term
induced on the
 brane, new phenomena appear. It was
observed \cite{dvali1}, that the localized matter fields on the
brane (which couple to bulk gravitons) can generate via quantum
loops a localized four-dimensional worldvolume kinetic term for
gravitons. That is to say, four-dimensional gravity is induced
from the bulk gravity to the brane worldvolume
 by the matter fields confined to the brane.

We will discuss static spherically symmetric solutions of
braneworlds with induced gravity \cite{pappa}, and we will present
an exterior Schwarzschild-$(A)dS_{4}$ solution which is matched to
a modified interior Oppenheimer-Volkoff solution. In this solution
the gravitational constant get corrected for very small matter
densities. The conventional solar system bounds of General
Relativity  set the crossover scale below the $TeV$ scale. The
above results were obtained by setting the electric part of the
Weyl tensor, $\textsf{E}_{\,\mu \nu}$ to zero.
\par
Then, we will generalize our study by including the non-local bulk
effects \cite{newkofinas}, as they are expressed by a
non-vanishing electric part of the Weyl tensor on the brane. By
choosing $g_{tt}=-g_{rr}^{-1}$, the system of equations on the
brane becomes closed and all the possible static black hole
solutions are found for these metrics. These solutions have
generic new terms which give extra attractive force compared to
the Newtonian -$(A)dS_{4}$ force, and represent the strong-gravity
corrections to the Schwarzschild-$(A)dS_{4}$ spacetime.

\section{Four-Dimensional Static Spherically Symmetric Solutions}
We consider a 3-dimensional brane $\Sigma$ embedded in a
5-dimensional spacetime. Capital Latin letters $A,B,...=0,1,...,4$
will denote full spacetime, lower Greek $\mu,\nu,...=0,1,...,3$
run over brane worldvolume. For convenience, we can quite
generally, choose a coordinate $y$ such that the hypersurface
$y=0$ coincides with the brane. The total action for the system is
taken to be:
\begin{eqnarray}
S&=&\frac{1}{2\kappa_{5}^{2}}\int\sqrt{-^{(5)}g}\,\,(^{(5)}R-2\Lambda_{5})d^{5}x+
\frac{1}{2\kappa_{4}^{2}}\int\sqrt{-^{(4)}g}\,\,(^{(4)}R-2\Lambda_{4})d^{4}x\nonumber
\\&+& \int\sqrt{-^{(5)}g}\,\,L_{5}^{mat}\,d^{5}x+
\int\sqrt{-^{(4)}g}\,\,L_{4}^{mat}\,d^{4}x. \label{action}
\end{eqnarray}
 We basically concern on the case with no fields in
the bulk, i.e. $^{(5)}T_{AB}=0$. From the dimensionful constants
$\kappa_{i}^{2}$, the Planck masses $M_{i}$ are defined as $
  \kappa_{i}^{2}=8\pi
G_{(i)}=M_{i}^{i-2}$
 Then, a distance scale
$r_{c}$ is defined as $r_{c}=M_{4}^{2}/ M_{5}^{3}$.

Varying (\ref{action}) with respect to the bulk metric $g_{AB}$,
and reducing the resulting equations to four dimensions
\cite{kofinas}, we get four-dimensional Einstein gravity, coupled
to a well-defined modified matter content. More explicitly, one
gets
\begin{equation}
^{(4)}G_{\nu}^{\mu}=\kappa_{4}^{2}\,^{(4)}T_{\nu}^{\mu}-\Big(\Lambda_{4}
+\frac{3}{2}\alpha^{2}\Big)\,\delta_{\nu}^{\mu}+
\alpha\Big(L_{\nu}^{\mu}+\frac{L}{2}\,\delta_{\nu}^{\mu}\Big)\,,
\label{einstein} \end{equation} where $\alpha\equiv 2/r_{c}$,
while the quantities $L^{\mu}_{\nu}$ are related to the matter
content of the theory through the equation
 \begin{equation}
L_{\lambda}^{\mu} L_{\nu}^{\lambda}-\frac{L^{2}}{4} \,
\delta_{\nu}^{\mu} = \mathcal{T}_{\nu}^{\mu} - \frac{1}{4} \Big{(}
3 \alpha^{2} + 2 \mathcal{T}_{\lambda}^{\lambda} \Big{)}
\delta_{\nu}^{\mu} \,, \label{lll} \end{equation} and $L\equiv
L^{\mu}_{\mu}$. The quantities $\mathcal{T}^{\mu}_{\nu}$ are given
by the expression \begin{eqnarray}
\mathcal{T}_{\nu}^{\mu}&=&\Big(\Lambda_{4}-\frac{1}{2}\,
\Lambda_{5}\Big)\delta_{\nu}^{\mu}
-\kappa_{4}^{2}\,^{(4)}T_{\nu}^{\mu}+\nonumber\\&&
+\frac{2}{3}\,\kappa_{5}^{2}\,\Big(\,^{(5)}\overline{T}\,_{\nu}^{\mu}
+\Big(\,^{(5)}\overline{T}\,_{y}^{y}-\frac{^{(5)}\overline{T}}{4}\Big)\,\delta_{\nu}^{\mu}\Big)
-\overline{\textsf{E}}^{\,\mu}_{\,\nu}\,.
 \label{energy}
 \end{eqnarray}
Bars over $^{(5)}T^{\mu}_{\nu}$ and the electric part\,
$\textsf{E}^{\,^{\mu}}_{\,\nu}$ of the Weyl tensor mean that the
quantities are evaluated at $y=0$. The presence of
$\overline{\textsf{E}}^{\,\mu}_{\,\nu}$
 makes the brane equations
(\ref{einstein}) not to be, in general, closed \cite{maeda}. Due
to the contracted Bianchi identities, the following differential
equations
 among $L^{\mu}_{\nu}$ arise from (\ref{einstein})
$L^{\mu}_{\nu ; \,\mu}+L_{;\, \nu}/{2}=0$. We are looking for
solutions of (\ref{einstein}) for spherically symmetric braneworld
metrics
\begin{equation} ds_{(4)}^{2}=-B(r)dt^{2}+A(r)dr^{2}+r^{2}(d
\theta^{2}+\sin ^{2} \theta d \phi^{2}). \label{spherical}
\end{equation} The matter content of the 3-universe is considered
to be a localized spherically symmetric perfect fluid.  We first
consider the case $\overline{\textsf{E}}^{\,\mu}_{\,\nu}=0$ as the
boundary condition of the propagation equations in the bulk space.
All the solutions outside a static localized matter distribution
were found. One of these is the Schwarzschild-$(A)dS_{4}$ metric
which is matched to a modified Oppenheimer-Volkoff interior. The
exterior solution was found to be
\begin{equation}
B_{>}(r)= \frac{1}{A_{>}(r)}=1-\frac{\gamma}{r}-\beta
r^{2}\,\,\,,\,\,\,r\geq R \,, \label{generalA>}
\end{equation}
where $\gamma$ is an integration constant and \begin{equation}
\beta=\frac{1}{3}\Lambda_{4}+\frac{1}{2}\alpha^{2}-
\frac{\alpha}{2\sqrt{3}}\sqrt{4\Lambda_{4}-2\Lambda_{5}+3\alpha^{2}}\,.
\label{beta1} \end{equation} Considering a uniform distribution
$\rho(r)=\rho_{o}=\frac{3M}{4\pi R^{3}}$ for the object, the
interior solution was found, and its matching to the above
exterior specified the integration constant $\gamma$. The result
is \begin{equation}
\frac{1}{A_{<}(r)}=1-(\beta+\frac{\gamma}{R^{3}})\,r^{2}\,\,\,,\,\,\,r
\leq R\,, \label{A<} \end{equation} \begin{equation}
B_{<}(r)=\frac{1-\frac{\gamma}{R}-\beta R^{2}}{\Big(1+\frac{4\pi
R^{3}}{3M}p(r)\Big)^{2}}\,\,\,,\,\,\,r\leq R\,, \label{B<}
\end{equation}
\begin{equation}
p(r)=-\rho_{o}\frac{\sqrt{1-(\beta+\frac{\gamma}{R^{3}})r^{2}}-
\sqrt{1-(\beta+\frac{\gamma}{R^{3}})R^{2}}}{\sqrt{1-(\beta+\frac{\gamma}{R^{3}})r^{2}}-
\omega \sqrt{1-(\beta+\frac{\gamma}{R^{3}})R^{2}}}\,,
\label{pressure} \end{equation} where \begin{equation}
\gamma=\frac{\kappa_{4}^{2}M}{4 \pi} + \frac{\alpha R^3
}{2\sqrt{3}}\sqrt{4\Lambda_{4}-2\Lambda_{5}+3\alpha^{2}} -
\frac{\alpha
R^3}{2\sqrt{3}}\sqrt{4\Lambda_{4}-2\Lambda_{5}+3\alpha^{2}+\frac{3\kappa_{4}^{2}M}{\pi
R^{3}}}\,, \label{gamma1} \end{equation} \begin{equation}
\omega^{-1}=1-\frac{2}{\kappa_{4}^{2}\rho_{o}}\left(\beta+\frac{\gamma}{R^3}\right)
\Big(1-\frac{\sqrt{3}\alpha}{\sqrt{4\Lambda_{4}-2\Lambda_{5}+3\alpha^{2}+4\kappa_{4}^{2}\rho_{o}}}
\Big)^{-1}\,. \label{omega} \end{equation} The parameters $\gamma$
and $\beta$ of the Schwarzschild-$(A)dS_{4}$ exterior solution
(\ref{generalA>}) can be constrained by solar system experiments.
The bounds obtained fix the crossover scale below the $TeV$ range.
The $\gamma$ parameter in the $1/r$ term modifies the Newton's
gravitational constant which, as it is seen from (\ref{gamma1}),
and for small matter it densities deviates significantly from its
conventional value. This is shown in Fig. 1.
%%%%%%%%%%%%%%%%%%%%%%%%%%%%%%%%%%%%%%%%%%%%%%
\begin{figure}[h!]
\centering
\includegraphics*[width=580pt, height=180pt]{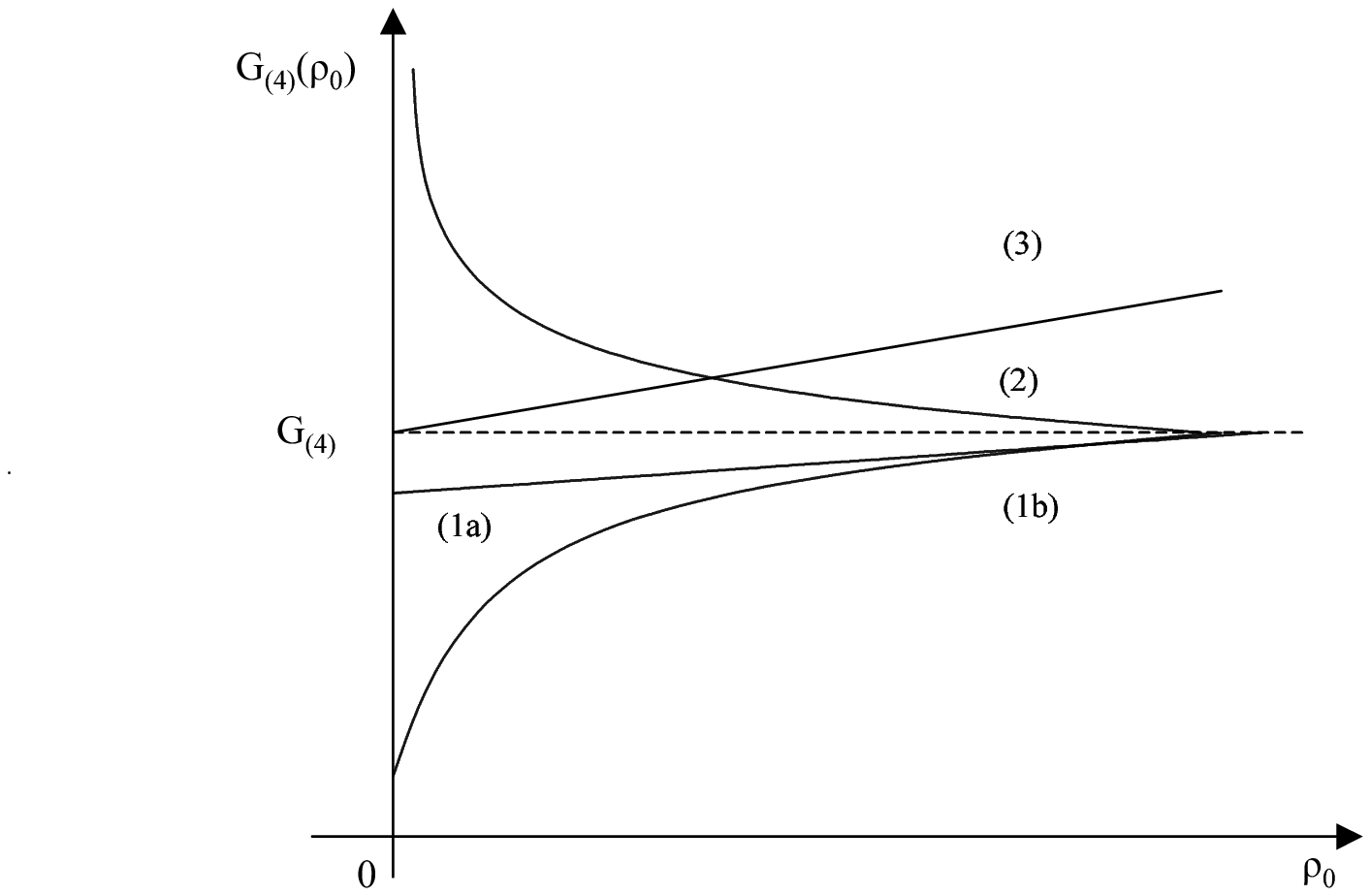}
%\includegraphics[scale=0.5]{plot2.ps}
%\caption
%\centerline{\hbox{\psfig{figure=figure.eps,height=10cm}}}
\caption {The $\rho_{0}$ dependence of Newton's constant in
various models:(1) The Newton's constant variation in our model;
If $4\Lambda_{4}-2\Lambda_{5}\gg 3\alpha^{2}$ then (1a), otherwise
(1b). (2) The behaviour of Newton's constant in another class of
our solutions. (3) The Newton's constant variation in RS model.}
\end{figure}
%%%%%%%%%%%%%%%%%%%%%%%%%%%%%%%%%%%%%%%%%%%%%%

So far we considered local corrections to the Einstein equations
on the brane. The presence of the electric part
$\overline{\textsf{E}}^{\,\mu}_{\,\nu}$ of the Weyl tensor  in
(\ref{energy}) indicates the 5D gravitational stresses, which are
known as massive KK modes of the graviton and this term cannot be
ignored.

The symmetric and traceless tensor
 $\overline{\textsf{E}}_{\mu \nu}$ is uniquely and
irreducibly
 decomposed as follows
 \begin{equation}
 \overline{\textsf{E}}_{\mu
\nu}=\mathcal{U}\Big(u_{\mu}u_{\nu}+\frac{1}{3}\textsf{h}_{\mu
\nu}\Big)+\mathcal{P}_{\mu \nu}+2\mathcal{Q}_{(\mu}u_{\nu)}\,,
\label{electric} \end{equation} where $\textsf{h}_{\mu \nu}=g_{\mu
\nu}+u_{\mu}u_{\nu}$ is the projection operator normal to
$u^{\mu}$, while  $\mathcal{U}$ is the non-local energy density,
$\mathcal{P}_{\mu \nu}$ the non-local anisotropic stress, and
$\mathcal{Q}_{\mu}$ the non-local energy flux on the brane. Static
spherical symmetry implies \cite{roys} that \begin{equation}
\mathcal{Q}_{\mu}=0\,\,\,\,,\,\,\,\,\mathcal{P}_{\mu
\nu}=\mathcal{P}(r)\Big(r_{\mu}r_{\nu}-\frac{1}{3}\textsf{h}_{\mu
\nu}\Big)\,, \label{sphericalrestriction} \end{equation} where
$r_{\mu}$ is the unit radial vector. We will discuss here one
class of solutions of (\ref{einstein}) with non-zero
$\textsf{E}_{\mu \nu}$. This is
\begin{eqnarray}
B=\frac{1}{A}&=&1-\frac{\gamma}{r}-\beta r^{2} \nonumber\\
&+&sg(\zeta)\,\frac{\delta}{r}\,\Big[
\frac{128}{105}\,_{1}F_{1}\Big(\frac{15}{8},\frac{23}{8};sg(\zeta)z\Big)\,z
\nonumber \\&+&
\frac{9}{8}\Big(\frac{1}{z}-sg(\zeta)\frac{8}{7}\Big)
\,e\,^{sg(\zeta)\,z}\Big]z^{\frac{7}{8}} \label{integration}
\end{eqnarray}
where $\delta$ and $\gamma$ are integration constants ( $\gamma$
is typically interpreted as $2G_{N}M$ with $M$ being the mass of
the point particle, $G_{N}$ the Newton's constant),
\begin{equation} \beta =\frac{1}{3} \Lambda_{4}+\frac{1}{2}
\alpha^{2}
\,\,\,\,\,\,,\,\,\,\,\,\,\zeta=\frac{\alpha^2}{9}(4\Lambda_{4}-2\Lambda_{5}+3\alpha^{2})\,,
\label{beta} \end{equation} and $
r=\Big(\frac{\delta}{\sqrt{|\zeta|}}\Big)^{\frac{1}{3}}z^{\frac{1}{8}}\,e\,^{sg(\zeta)\,z/3}\,.
\label{laf}$ The electric components of the Weyl tensor are given
by the equation, $\mathcal{P}=-2\mathcal{U}=9\zeta/(2\alpha^2)$.
Comparing the solution (\ref{integration}) with the solution
(\ref{generalA>}) having $\overline{\textsf{E}}^{\mu}_{\nu}=0$, we
note that there is a new term, besides the conventional Newtonian
and $(A)dS_{4}$ terms, which carries the information of the
gravitational field in the bulk. This new term modifies the
Newton's law for small and large distances as it is shown in Fig.
2.
\begin{figure}[h!]
\centering
%\hspace{0.1cm}%
\begin{tabular}{cc}
\includegraphics*[width=150pt, height=120pt]{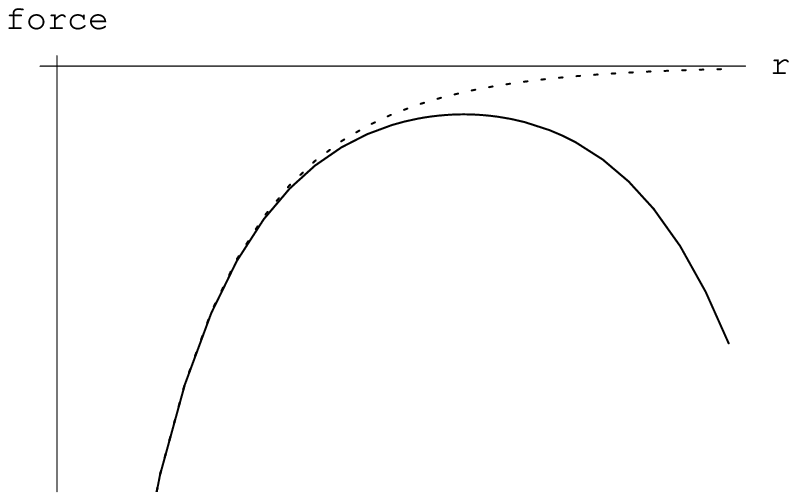}&%
%\hspace{0.1cm}% \\
\includegraphics*[width=150pt, height=120pt]{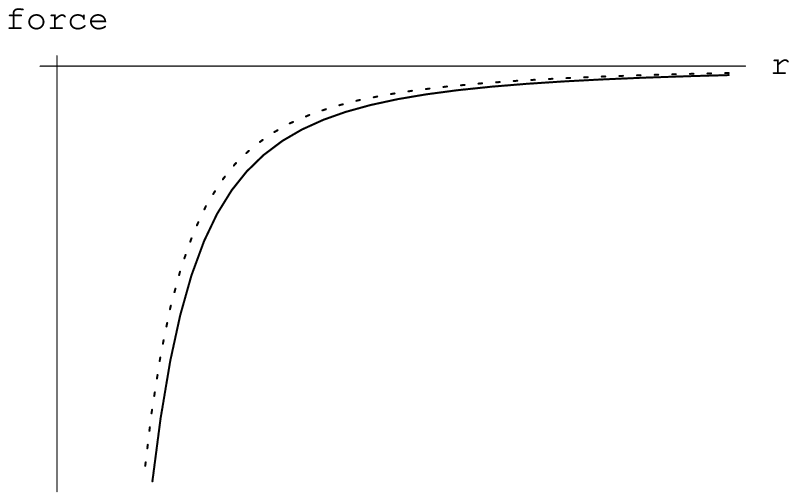}\\%
(a): $\zeta >0$, $\beta=0$&(b):  $\zeta <0$, $\beta=0$
\end{tabular}
 \caption{
Dotted lines represent the Newtonian force. Continuous lines
represent the total force, i.e. the sum of the Newtonian and the
new force.}
\end{figure}
\section{Conclusions}
We presented new black hole solutions on the brane with induced
gravity. Solving the full theory taking under consideration the
non-local bulk effects,  we found static spherically symmetric
solutions, which represent strong gravity corrections to the
Schwarzschild-$(A)dS_{4}$ black hole solutions. Their
characteristic is that they predict a new attractive force. There
are classes of solutions with increasing $r$, where this
attractive force combined with the Newtonian one, results to a net
force which decreases slower than the Newton's force. In another
class of solutions with decreasing $r$, the new force starts to
deviate from the Newton's force at small distances, indicating
that at submillimeter scale we could have testable deviations from
the Newtonian law. We have also found modifications to Newton's
law as a result of a change of the Newton's constant due to the
finite interior of the rigid object.

\end{article}
\end{document}